\documentclass[aps,prb,twocolumn,groupedaddress,showpacs]{revtex4}
\usepackage{graphicx}
\usepackage{epstopdf}
\usepackage{amssymb, amsmath}
\usepackage{multirow}
\usepackage{dcolumn}
\usepackage{bm}
\usepackage{subfigure}
\usepackage{color}
\usepackage{ulem}

\usepackage[final]{hyperref}
\hypersetup{
        colorlinks=true,       
        linkcolor=blue,          
        citecolor=blue,        
        filecolor=magenta,      
        urlcolor=blue
        }

\newcommand\ME[3]      {\langle{{#1}}|{{#2}}|{{#3}}\rangle} 
\newcommand\ket[1]     {|{{#1}}\rangle}
\newcommand\bra[1]     {\langle{{#1}}|}
\newcommand\braket[2]  {\langle{{#1}}|{{#2}}\rangle}

\newcommand\PsiGS      {\Psi_0}
\newcommand\Dc[1]      {c_{{#1}}^{}}
\newcommand\Cc[1]      {c_{{#1}}^\dagger}
\newcommand\Half       {\frac{1}{2}}
\newcommand\Gvec       {\mathbf{G}}
\newcommand\Qvec       {\mathbf{Q}}
\newcommand\kvec       {\mathbf{k}}

\newcommand\rvec       {\mathbf{r}}

\newcommand\Hop        {{\hat{H}}}
\newcommand\Kop        {{\hat{K}}}

\newcommand\Vop        {{\hat{V}}}

\newcommand\Veiop      {\hat{V}_{\textrm{ei}}}
\newcommand\Veeop      {\hat{V}_{\textrm{ee}}}

\newcommand\VL         {V_{\textrm{L}}}
\newcommand\VaL         {V_{\alpha,\textrm{L}}}
\newcommand\VNL        {V_{\textrm{NL}}}

\newcommand\Vii        {V_{\textrm{II}}}
\newcommand\HSop       {{\hat{b}}}

\newcommand\ABINIT     {{\footnotesize{ABINIT}}}
\newcommand\ELK    {{\footnotesize{ELK}}}

\newcommand\eql[2] 
{
\begin{equation}\label{#1}
\begin{split}
#2
\end{split}
\end{equation}
}

\newcommand\eqsl[1]                            
{
\begin{align}
#1
\end{align}
}

\newcommand\COMMENTED[1] {}

\begin{document}
 
\title{Auxiliary-field quantum Monte Carlo calculations with \\ multiple-projector pseudopotentials}
 
\author{Fengjie Ma}
\altaffiliation[Present affiliation: ]
{Center for Advanced Quantum Studies and Department of Physics, Beijing Normal University, Beijing 100875, China}
\email{fengjie.ma@bnu.edu.cn}
\affiliation{Department of Physics, College of William and Mary,
Williamsburg, Virginia 23187, USA}

\author{Shiwei Zhang}
\affiliation{Department of Physics, College of William and Mary,
Williamsburg, Virginia 23187, USA}

\author{Henry Krakauer}
\affiliation{Department of Physics, College of William and Mary,
Williamsburg, Virginia 23187, USA}

\begin{abstract}

We have implemented recently developed multiple-projector pseudopotentials into the planewave based auxiliary-field quantum Monte Carlo (pw-AFQMC) method. Multiple-projector pseudopotentials can yield smaller planewave cut-offs while maintaining or improving transferability. 
This reduces the computational cost of pw-AFQMC, increasing its reach to
larger and more complicated systems. 
We discuss the use of non-local pseudopotentials in the separable Kleinman-Bylander form, 
and the implementation in pw-AFQMC of the multiple-projector 
optimized norm-conserving pseudopotential ONCVPSP of Hamann.
The accuracy of the method is first demonstrated by equation-of-state  
calculations of the ionic insulator NaCl and more strongly correlated metal Cu. 
The method is then applied to calibrate the accuracy of density functional theory (DFT) predictions of the phase stability of recently discovered high temperature and pressure superconducting sulfur hydride systems. 
We find that DFT results are in good agreement with pw-AFQMC, due to 
near cancellation of electron-electron correlation effects between different structures.

\end{abstract}



\maketitle

\section{Introduction}

The search for new materials and their development has increasingly relied on theoretical modeling. Methods based on density functional theory (DFT) are efficient and powerful, but their predictions can break down in a number of instances. Examples range from strongly
correlated materials, such as transition metal systems, to bond stretching or bond breaking in otherwise moderately correlated systems. Explicit many-body methods, which avoid the mean-field-like 
approximations used in standard DFT calculations, are needed in these cases. Quantum Monte Carlo
(QMC) calculations have become increasingly important in this regard, because of their accuracy and favorable scaling (as a low-order polynomial of system size, similar to DFT, but with larger prefactor) compared to traditional wave function based correlated methods. 
Routine applications of QMC calculations in extended systems still face major challenges, however. 

In diffusion QMC (DMC) \cite{QMCRMP2001} and pw-AFQMC, \cite{AFQMC,Suewattana2007,AFQMCFC,AFQMCdownfolding}  
pseudopotentials are usually used, except for some DMC calculations with the lightest elements.  
Pseudopotentials 
remove the chemically inactive core electrons, reducing the number of electrons that must be explicitly correlated
and greatly reducing the computational cost. 
Non-local norm-conserving pseudopotentials (NCPP) are typically used in QMC.
The NCPPs are usually constructed from mean-field DFT of Hartree-Fock (HF) calculations.
While computationally expedient, the transferability of NCPPs is a key issue, and the neglected core-core and core-valence correlation effects may need
to be considered. Even setting these many-body effects aside, 
transferability errors from NCPPs in QMC calculations can be significant. 
In DMC, moreover, the non-locality of NCPPs is handled with an additional locality approximation,
whose accuracy depends on the quality of the trial wave function. \cite{DMClocality} 
The overall NCPP error can be significant compared to  
errors from the fixed-node 
approximation, \cite{Hennig2010,Sorella2011,DMC-psp-benchmark2016} 
which is used to control the Fermion sign problem. In pw-AFQMC, non-local NCPPs can be used without additional 
approximations,\cite{AFQMC,Suewattana2007} but transferability errors can still be a problem, unless the NCPPs are made very hard, \cite{Suewattana2007,AFQMCSi,AFQMCFC,AFQMCdownfolding} 
which requires large planewave cutoffs  
and increases the computational cost.

Pseudopotentials are based on the frozen-core approximation, but contain an additional layer of approximation. 
Frozen-core calculations are common in quantum chemistry applications, where the core orbitals are frozen at the mean-field level derived from the {\it target} system. 
Pseudopotentials are usually constructed for a reference {\it atomic} configuration
 and then used in many target systems. 
The accuracy (transferability) of the PP across many target systems must then
be determined {\it a posteriori}. 
In addition to being norm-conserving, most NCPPs used in QMC calculations are of single-projector type (one per angular momentum channel), which can further contribute to transferability errors.

Recently, Hamann proposed a multiple-projector pseudopotential,\cite{ONCVPSP} based on Vanderbilt's norm-conserving construction \cite{USPP} and optimized with the  Rappe-Rabe-Kaxiras-Joannopoulos 
pseudization scheme. \cite{RRKJ} 
The resulting pseudopotential, referred to as \mbox{ONCVPSP} by Hamann, was shown to have accuracy comparable to all-electron (AE) and ultrasoft pseudopotentials \cite{USPP} (USPP) in DFT calculations, with moderate planewave energy cutoffs. 
Schlipf and Gygi  \cite{Schlipf-Gygi-ONCVPSP} recently presented a set of automatically constructed Hamann ONCVPSPs for most of the periodic table. These were shown to be in good agreement with the all-electron results in DFT, often with cutoffs of only about 40 Ry. \cite{Schlipf-Gygi-ONCVPSP,DFT-Reproducibility}
The ONCVPSP is of 
separable Kleinman-Bylander type, \cite{KB} similar to NCPPs widely used in planewave DFT calculations and also in pw-AFQMC calculations. 
Since the treatment of one-particle Hamiltonian terms in  pw-AFQMC is closely related to that in planewave DFT,
the implementation of ONCVPSP into our pw-AFQMC code is  
straightforward. 

In this paper, we show that the use of multiple-projector ONCVPSP 
can greatly reduce the planewave basis size in pw-AFQMC many-body calculations, while maintaining good accuracy. 
This results in significant reductions of  computational cost, both by reducing the computing time for each
step in the random walks and, at the same time, by reducing QMC statistical variance, due to the reduced number
of AFQMC auxiliary fields.
To test the new capability with multiple-projector ONCVPSPs, we carry out pw-AFQMC calculations of the equation-of-state in  the insulator NaCl and the transition metal solid Cu. We then study the 
high-pressure superconducting system H$_3$S, to calibrate DFT predictions of phase stabilities. 
Finally we discuss the performance of the DFT- or HF-generated pseudopotentials in many-body calculations and the difference from their use in DFT calculations.

The reminder of the paper is organized as follows. Section~\ref{sec:Method} reviews AFQMC with a planewave basis and pseudopotentials, and discusses the implementation of multiple-projector separable pseudopotentials
in pw-AFQMC.
Section~\ref{sec:Applications} presents applications of the method.
Additional transferability issues and other aspects of \mbox{ONCVPSP} for many-body applications are discussed in Section~\ref{sec:Discussion}.
We then conclude with some general remarks in Section~\ref{sec:Summary}.

\section{pw-AFQMC methodology}
\label{sec:Method}

To set the context for the implementation of multiple-projector NCPPs  in pw-AFQMC, we briefly review pertinent aspects of the
formalism in this section. For more details about the pw-AFQMC method, see Refs.~[\onlinecite{AFQMC,Suewattana2007,AFQMCSi}].

\subsection{Hamiltonian}

The electronic Hamiltonian within the Born-Oppernheimer approximation is,
\begin{equation}
\label{eq:H0}
    \Hop
  = \Kop + \Veeop + \Veiop + \Vii
    \,,
\end{equation}
where $\Kop$, $\Veeop$, $\Veiop$, and $\Vii$ are, respectively, the kinetic energy and electron-electron, electron-ion, 
and classical Coulomb ion-ion\cite{Yin_1982} 
interactions. The pseudopotential contributions appear in the electron-ion interaction  $\Veiop$.
With periodic boundary conditions and a planewave basis, 
\begin{equation}
   \braket{\rvec}{\Gvec}
   \equiv
   \ME{{\rvec}}{\,\Cc{\Gvec}}{0}
 = \frac{1}{{\sqrt \Omega}}\exp (i{\Gvec} \cdot {\rvec})
   \, ,
\end{equation}
the terms in Eq.~(\ref{eq:H0})  can be expressed in second quantized form as
\begin{subequations}
\begin{align}
\label{H_terms_K}
    \Kop
&  = \Half \sum_{\Gvec,s} {G^2}\, \Cc{\Gvec,s} \Dc{\Gvec,s}
    \,,
\\
    \Veeop
\label{H_terms_V2}    
& = \Half N \xi
  + \frac{1}{2\Omega} \sum_{\Qvec \neq \mathbf{0}}
    \frac{4\pi}{Q^2}\,\hat{\rho}^\dagger(\Qvec) \hat{\rho}(\Qvec) \nonumber
\\
& - \,
    \frac{1}{2\Omega} \sum_s \sum_{\Gvec,\Gvec'}
    \frac{4\pi}{\left|\Gvec - \Gvec'\right|^2}
    \Cc{\Gvec,s} \Dc{\Gvec,s}
    \,,
\\
    \Veiop
& = \Half
    \sum_{\Qvec \neq \mathbf{0}} \VL(\Qvec)
    \left[
        \hat{\rho}({\Qvec}) +
        \hat{\rho}^\dag({\Qvec})
    \right]  \nonumber \\
& + \sum_{\Gvec,\Gvec'}
    \VNL({\Gvec}, {\Gvec'}) \Cc{\Gvec} \Dc{\Gvec'}
   + N\VL(\mathbf{0})
    \,.
\label{H_terms_PSP}    
\end{align}
\end{subequations}
Here, $\Cc{\Gvec}$ ($\Dc{\Gvec}$) is a creation (destruction) operator, $\Omega$ is the volume of the simulation cell, and ${\Gvec}$ is a reciprocal lattice vector, 
\mbox{$\Qvec = \Gvec' - \Gvec$}, 
$s$ is the electron spin, and $N$ is the number of electrons in the simulation cell.
Both ${\Gvec}$ and ${\Gvec'}$ belong to the  planewave basis set $\{{\Gvec}\}$ whose size 
is controlled by the planewave kinetic energy cut-off \mbox{$E_\mathrm{cut}\ge \left|{\Gvec}\right|^2/2$}.
(When twist-averaged boundary conditions are used, ${\Gvec}$ is replaced by ${\kvec+\Gvec}$, where $\kvec$ is within the first Brillouin zone.)
The constant $\xi$ gives the self-interaction of an electron with its periodic images. \cite{Fraser1996}
The one-body density operator $\hat{\rho}({\Qvec})$ is given by
\begin{equation}
    \label{rho}
    \hat{\rho}({\Qvec})
    \equiv
    \sum_{\Gvec,s}
    \Cc{{\Gvec + \Qvec},s} \Dc{{\Gvec},s}
    \,
    \theta
    \left(
        E_\mathrm{cut} - \left|{\Gvec+\Qvec}\right|^2/2
    \right)
    ,
\end{equation}
where the step function $ \theta$ ensures that $({\Gvec + \Qvec})$, like $\Gvec$, falls within 
the planewave basis set.

The local and non-local parts of the pseudopotential are defined by the
planewave matrix elements
$\VL({\Qvec})$ and $\VNL ({\Gvec},{\Gvec'})$, respectively, which are discussed in more detail in Section \ref{sec:PSP}.

\subsection{Ground state projection}
\label{sec:GSprojection}

AFQMC uses iterative imaginary-time projection to obtain the ground state $\ket{\PsiGS}$ from a trial wave function $\ket{\Psi_T}$ 
(often just a single Slater determinant):
\eql{eq:gs-proj-beta}
{
    e^{-\beta \Hop}
    \ket{\Psi_T}
  \rightarrow \ket{\PsiGS} ~~~~(\beta \rightarrow \infty)
    \, ,
}
where $\braket{\Psi_T}{\PsiGS} \neq 0$ is assumed. 
The projection 
is implemented as random walks in the space of Slater determinants. 
A key point in implementing this is the observation that a one-body propagator
acting on a Slater determinant simply yields another Slater determinant. 
The AFQMC procedure is therefore to separate the propagator in Eq.~(\ref{eq:gs-proj-beta})
into one- and two-body propagators. 
This motivates the 
introduction of a small imaginary-time step $\Delta \tau$:
\eql{eq:gs-proj}
{
    e^{-\Delta \tau \Hop}
    e^{-\Delta \tau \Hop}
    \cdots
    e^{-\Delta \tau \Hop}
    \ket{\Psi_T}
  \rightarrow \ket{\PsiGS}
    \,.
}
A Trotter-Suzuki
decomposition \cite{Trotter1959,Suzuki1976} then achieves the desired separation,
\eql{eq:Trotter}
{
    e^{-\Delta \tau \Hop}
& =e^{-\frac{1}{2}\Delta \tau \Hop^{(1)}}
    e^{-\Delta \tau \Hop^{(2)}}    
    e^{-\frac{1}{2}\Delta \tau \Hop^{(1)}}
 + \, O(\Delta \tau^3)
    \, ,
}
where  
$\Hop^{(1)}$ and $\Hop^{(2)}$ are the one- and two-body parts of the Hamiltonian in Eq.~(\ref{eq:H0}),
with  $\Hop^{(2)}=
1/(2\Omega) \sum_{\Qvec \neq \mathbf{0}}
    \frac{4\pi}{Q^2}\,\hat{\rho}^\dagger(\Qvec) \hat{\rho}(\Qvec)$ and  $\Hop^{(1)}$
    denoting the remaining terms in Eq.~(\ref{H_terms_V2}) and the
    collection from Eqs.~(\ref{H_terms_K}) and (\ref{H_terms_PSP}).

A Hubbard-Stratonovich transformation \cite{Hubbard,Stratonovich} allows one to express two-body propagators 
as a high-dimensional integral over auxiliary fields $\{\sigma_i\}$ of one-body propagators:
\eql{eq:HS-xform0}
{
&   \exp{\left({-\Half \Delta\tau \sum_i\lambda_i \HSop_i^2}\right)}
\\
&\quad\;
    =
    \int \left( \prod_i \frac{d\sigma_i}{\sqrt{2\pi}} \right)
    \exp{\left[\sum_i \left(-\Half \sigma_i^2
    +\sigma_i \sqrt{-\Delta\tau\lambda_i} \, \HSop_i \right)
    \right]}\, ,
}
where the $\hat{b}_i$ are any one-body operators.
Applying this to $e^{-\Delta\tau \Hop^{(2)}}$
we have
\eql{eq:HS-xform}
{
    e^{-\Delta\tau\Hop^{(2)}}
  = \left(\frac{1}{\sqrt{2\pi}}\right)^{D_\sigma}
    \!
    \int d\bm{\sigma}  \,
    e^{-(1/2) \bm{\sigma} \cdot \bm{\sigma}}
    e^{\sqrt{\Delta\tau}\, \bm{\sigma} \cdot \hat{\mathbf{v}} }
    \, ,
}
where we have introduced the vector of auxiliary fields $\bm{\sigma} \equiv \{\sigma_i\}$,
whose 
dimension, $D_\sigma$, is given by the number of
possible $\Qvec$-vectors. The operators $\hat{\mathbf{v}} \equiv \{ \sqrt{-\lambda_i}\,\HSop_i \}$
are given by  
linear combinations of $\hat{\rho}^\dagger(\Qvec)$ and $\hat{\rho}(\Qvec)$. 
\cite{AFQMC,Suewattana2007}

Our focus in this paper is on the choice of pseudopotentials, 
which appear only in the one-body propagator $e^{-\frac{1}{2}\Delta \tau \Hop^{(1)}}$.
The handling of the two-body propagator and the implementation of the AFQMC phaseless approximation are unchanged
from previous applications and can be found in  Refs.~[\onlinecite{AFQMC,Suewattana2007,AFQMCSi}].

The overall computing cost in QMC depends not only on the computer time to execute a single time step for each random walker, 
but also on the statistical variance, which controls the size of the Monte Carlo sampling required to achieve a targeted statistical uncertainty
 (the QMC efficiency).\cite{AFQMCdownfolding}
The computing cost to execute a single imaginary-time step 
[Eq.~(\ref{eq:gs-proj})] in pw-AFQMC is proportional to $M\ln(M)$, 
where $M \propto E_\mathrm{cut}^{3/2}$ is the number of planewaves. [The overall scaling is $N^2M\ln(M)$, where $N$ is the number
of electrons in the simulation cell.] The statistical variance depends on the
number of auxiliary fields $D_\sigma \propto 8\,E_\mathrm{cut}^{3/2}$. 
Reducing  $E_\mathrm{cut}$ can therefore both reduce  the computing time for each
step in the random walk and increase the QMC efficiency.
Convergence with respect to $E_\mathrm{cut}$ is controlled by the pseudopotential hardness,\cite{AFQMC,Suewattana2007} so that soft accurate pseudopotentials can potentially lead
to major improvements in pw-AFQMC.

\subsection{Pseudopotential}
\label{sec:PSP}

The pseudopotential appears in the electron-ion interaction, \mbox{$\Veiop= \Vop_L + \Vop_{NL}$} in Eq.~(\ref{H_terms_PSP}).
In second-quantized form, the pseudopotential's action is safely isolated in the planewave matrix elements of its local [$\VL({\Qvec})$] 
and non-local [$\VNL ({\Gvec},{\Gvec'})$] parts, exactly as in DFT planewave methods.\cite{Yin_1982} 
Non-local potentials thus present no difficulties in AFQMC (unlike
in the real-space-based DMC method \cite{DMClocality}).
The planewave matrix elements of the local part of the pseudopotential are given by
\begin{equation}
\label{eq:psp-loc}
    \VL(\Qvec)
  = \frac{1}{\Omega} \sum_\alpha e^{-i {\Qvec} \cdot \mathbf{d_{\alpha}}} \VaL(\Qvec)
    \,,
\end{equation}
where $\mathbf{d_{\alpha}}$ is the 
position of atom $\alpha$ in the simulation cell, and $\VaL(\Qvec)$ is the Fourier transform of the (spherical) local part of the 
atomic pseudopotential. 
For single-projector NCPPs, the non-local part of the atomic pseudopotential is expressed by the separable Kleinman-Bylander \cite{KB} form,
\begin{equation}
\label{eq:psp-kb}
\hat{V}_{\alpha l,\textrm{NL}}
  = \sum_{m=-l}^{l}
       \frac{\ket{V^{\rm ps}_{\alpha,l} \varphi_{\alpha,l} Y_{lm}} \bra{Y_{lm} \varphi_{\alpha,l} V^{\rm ps}_{\alpha,l}}}
          {\ME{Y_{lm} \varphi_{\alpha,l}}{V^{\rm ps}_{\alpha,l}}{\varphi_{\alpha,l} Y_{lm}}}
    \,,
\end{equation}
where, for each partial wave ({\it e.g.}, $l=0, 1, 2$ for $3d$ transition elements), there is only one projector. 
(The pseudopotential $V^{\rm ps}_{\alpha,l}$ and pseudo-orbital $\varphi_{\alpha,l}$ are both functions of 
 radial distance $r$ only, and  $Y_{lm}$ is the usual spherical harmonic function.)
The matrix elements $\VNL ({\Gvec},{\Gvec'})$ of the non-local part of the pseudopotential can then be expressed in a separable form as,
\begin{equation}
\label{eq:psp-nloc}
    \VNL ({\Gvec},{\Gvec'})
  = \sum_{j\in{\{ \alpha,l,m} \}}
    \frac{1}{\eta_j}
    F^{*}_j (\Gvec) F_j(\Gvec')
    \,,
\end{equation}
where $\eta_j= \ME{Y_{lm} \varphi_{\alpha,l}}{V^{\rm ps}_{\alpha,l}}{\varphi_{\alpha,l} Y_{lm}}$, and
\begin{equation}
\label{eq:psp-Fjq}
    F_j(\Gvec)
  = \frac{4\pi}{\sqrt{\Omega}}
      e^{i \Gvec \cdot \mathbf{d_{\alpha}}} f_{\alpha,l}(\Gvec) Y^{*}_{lm}(\hat\Gvec)
    \,,
\end{equation}
where 
$f_{\alpha,l}(\Gvec)$ is obtained 
from the Bessel transform of the projector $\ket{V^{\rm ps}_{\alpha,l} \varphi_{\alpha,l} Y_{lm}}$.
The separable form of $\VNL ({\Gvec},{\Gvec'})$ greatly simplifies and speeds up the use of the NCPPs, 
just as in DFT methods.

Equation\,(\ref{eq:psp-kb}) can be abbreviated as $\hat{V}_{\alpha l,\textrm{NL}} \equiv \frac{|\chi_1\rangle\langle\chi_1|}{b_1}$, in which $b_1$ is the overlap between pseudo-wavefunction $\phi^{\rm ps}_{\alpha,l}$ with constructed projector $|\chi\rangle ={V^{\rm ps}_{\alpha,l} |\phi^{\rm ps}_{\alpha,l}\rangle} $.
Having only one projector for each partial wave $l$ limits the energy range over which
an NCPP can reproduce the scattering properties of the
all-electron potential, which reduces its transferability. 
Hamann generalized Eq.\,(\ref{eq:psp-kb}) for optimized multiple projectors (in practice, implemented for two projectors per
partial wave). \cite{ONCVPSP}
Written in a diagonal representation, the multiple-projector pseudopotential
can be compactly expressed for atom $\alpha$ and partial wave $l$ as \cite{ONCVPSP}
\begin{equation}
\label{eq:psp-ONCV}
\hat{V}_{\alpha l,\textrm{NL}}
  = \sum_{i=1}^{2}
    \frac{\ket{\chi_i} \bra{\chi_i}}
         {b_i}  
    \, .
\end{equation}
Implementing the ONCVPSP in this form requires only minor modifications in pw-AFQMC, compared to Eq.\,(\ref{eq:psp-kb}). 
The extended energy range over which the scattering properties are reproduced often allows smaller planewave $E_\mathrm{cut}$ with
excellent transferability properties. \cite{ONCVPSP}

\section{Applications}
\label{sec:Applications}

We describe applications of pw-AFQMC with \mbox{ONCVPSP} in three systems, the ionic insulator NaCl, the strongly correlated metal Cu, and the recently discovered sulfur hydride high-T$_c$ and high-pressure superconductors.

\subsection{Ionic insulator: NaCl}

\begin{figure}
\begin{center}
\includegraphics[width=0.4\textwidth]{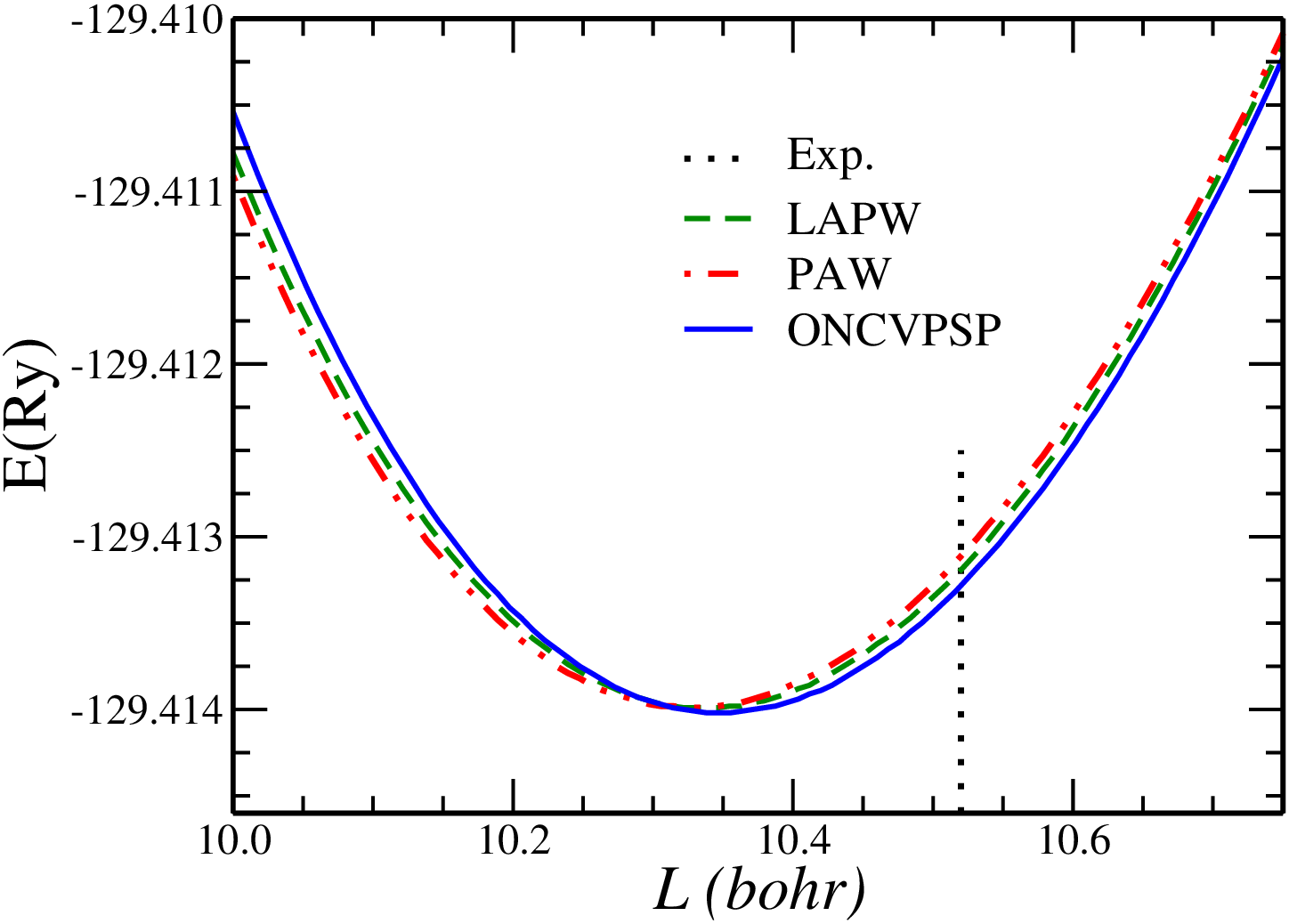}
\caption{\label{NaCl_psps} (Color online) NaCl DFT/LDA calculated EOS curves (fits to Murnaghan's equation \cite{MurnaghanEOS}), comparing all-electron LAPW (green dashed line), PAW (red dot-dashed line) and ONCVPSP (blue solid line). Curves are shifted to have the same minimum energy. The experimental lattice constant is indicated by the black dotted vertical line. 
}
\end{center}
\end{figure}

NaCl is a typical ionic compound, which crystallizes in the fcc structure.
Due to the large overlap of Na valence $3s$ electron and semicore $2s$ and $2p$ states, care must be used in the choice of pseudopotentials. 
Relaxation of the  semicore states can significantly affect valence electron and hence material properties. 
Neglecting these effects, for example, by treating the $2s$ and $2p$ electrons as core states will give \mbox{$\sim 10\%$} underestimation of the lattice constant and \mbox{$45\%$} overestimation of the bulk modulus of Na in DFT 
calculations. Pseudizing instead the $2s$ and $2p$ states greatly reduces the discrepancy in lattice constant,
to \mbox{$\sim 1.6\%$} 
using the local density approximation (LDA) exchange-correlation functional.
Calculated equation of states (EOS) with LDA 
are shown in Fig.~\ref{NaCl_psps}.  
ONCVPSP results are compared to those from the 
all-electron linearized augmented planewave (LAPW) and the projector augmented wave (PAW) methods, 
using \ELK \cite{elk} and  \ABINIT \cite{abinit}, respectively.
ONCVPSPs were generated with Hamann's open source pseudopotential code. \cite{oncvcode}
Both \mbox{ONCVPSP} and PAW results are in excellent agreement with LAPW. 
The agreement can be further quantified, using the $\Delta$ factor, which was recently introduced by Lejaeghere {\it et al.}\cite{Lejaeghere}
for comparing two EOS curves, $E_{1}(V)$ and $E_{2}(V)$.
Aligning the minimum energies, 
the definition of $\Delta$ is:
\begin{equation}
\Delta=\sqrt{\frac{\int{[E_{2}(V)-E_{1}(V)]^2 dV}}{\Delta{V}}}
\end{equation}
for a volume range $\Delta V$. (A typical choice of $\Delta V$ is $\pm6\%$ around the equilibrium volume.)
The $\Delta$ factors are $0.89$\,meV and $0.79$\,meV for \mbox{ONCVPSP} and PAW calculations, respectively.  
The Na and Cl multiple-projector  
ONCVPSP pseudopotentials required
kinetic energy 
cut-offs of only \mbox{$E_\mathrm{cut}=40$\,Ry}, much softer than 
for a single-projector norm-conserving pseudopotentials,
which would have required \mbox{$E_\mathrm{cut}=100$\,Ry}.

\begin{figure}
\begin{center}
\includegraphics[width=0.4\textwidth]{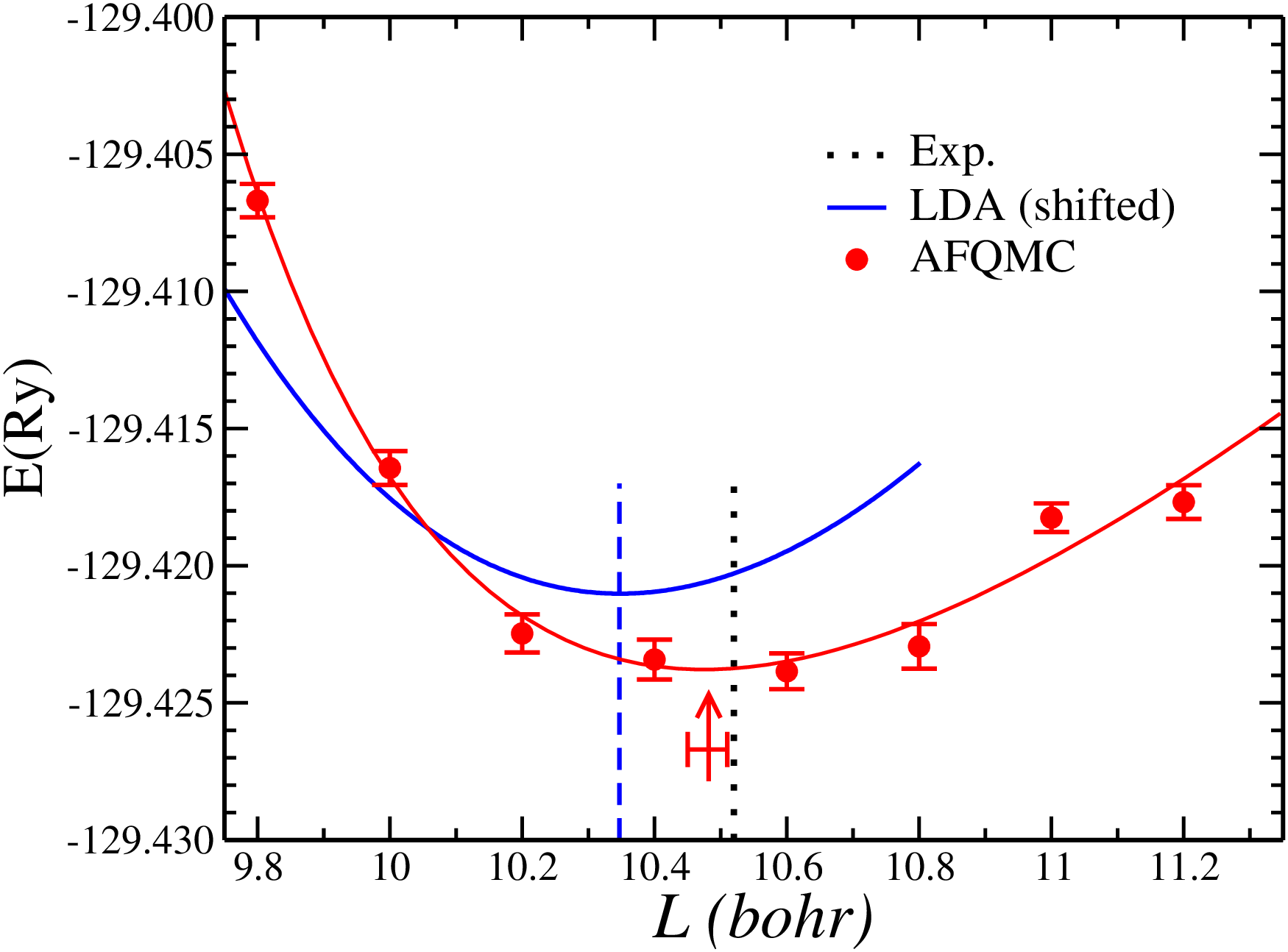}
\caption{\label{NaCl_EOS} (Color online) 
NaCl EOS calculated from pw-AFQMC (filled red circles with statistical error bars) using the same ONCVPSPs as
in Fig.~\ref{NaCl_psps}. 
For comparison, the DFT EOS in Fig.~\ref{NaCl_psps} is reproduced  (blue solid line, energy shifted for convenient display). 
The vertical blue dashed and black dotted lines indicate the DFT and experimental equilibrium lattice constants $a_0$, respectively.
The pw-AFQMC calculated $a_0$ 
is indicated by the vertical red arrow with 
horizontal error bar indicating the uncertainty  from a fit of the statistical data to Murnaghan's equation. \cite{MurnaghanEOS}
}
\end{center}
\end{figure}

Figure \ref{NaCl_EOS} shows the calculated pw-AFQMC NaCl EOS.
A  four-formula cubic simulation cell was used
with twist boundary condition corresponding to the $L$ special $\mathbf{k}$-point ($0.5$, $0.5$, $0.5$);
one- and two-body finite-size errors 
were reduced, using a post-processing finite-size correction scheme. \cite{KZKFS,KZKFSspin} 
Here and throughout the rest of this paper, 
Trotter errors from Eq.~(\ref{eq:Trotter}) are removed by either extrapolation to the 
 \mbox{$\Delta \tau=0$} limit or choosing sufficiently small  time-step values.
Our pw-AFQMC calculations used LDA-generated trial wave functions.  
The discrepancy of DFT/LDA with experiment is essentially eliminated by the many-body calculations.
The equilibrium lattice constant and bulk modulus calculated from pw-AFQMC,
\mbox{$a_0=10.48(3)$\,bohr} and  \mbox{$B_0=26(2)$\,GPa}, are in excellent agreement
with the experimental values,  \mbox{$a_0=10.52$\,bohr} and  \mbox{$B_0=26.6$\,GPa}. \cite{ExpNaClCu}

\subsection{Transition metal: fcc Cu}

Transition metal materials have played a central role in the study of strongly correlated physics, and
copper based systems have especially attracted a great deal of attention. \cite{RMPPickett} 
{\it Ab-initio} many-body calculations for transition metal systems have been very limited, \cite{DMCCuprate} and most previous calculations
have relied on DFT or related approaches. 
In this subsection, we present many-body pw-AFQMC results 
on fcc copper,  a prototypical correlated metal.

\begin{figure}
\begin{center}
\includegraphics[width=0.4\textwidth]{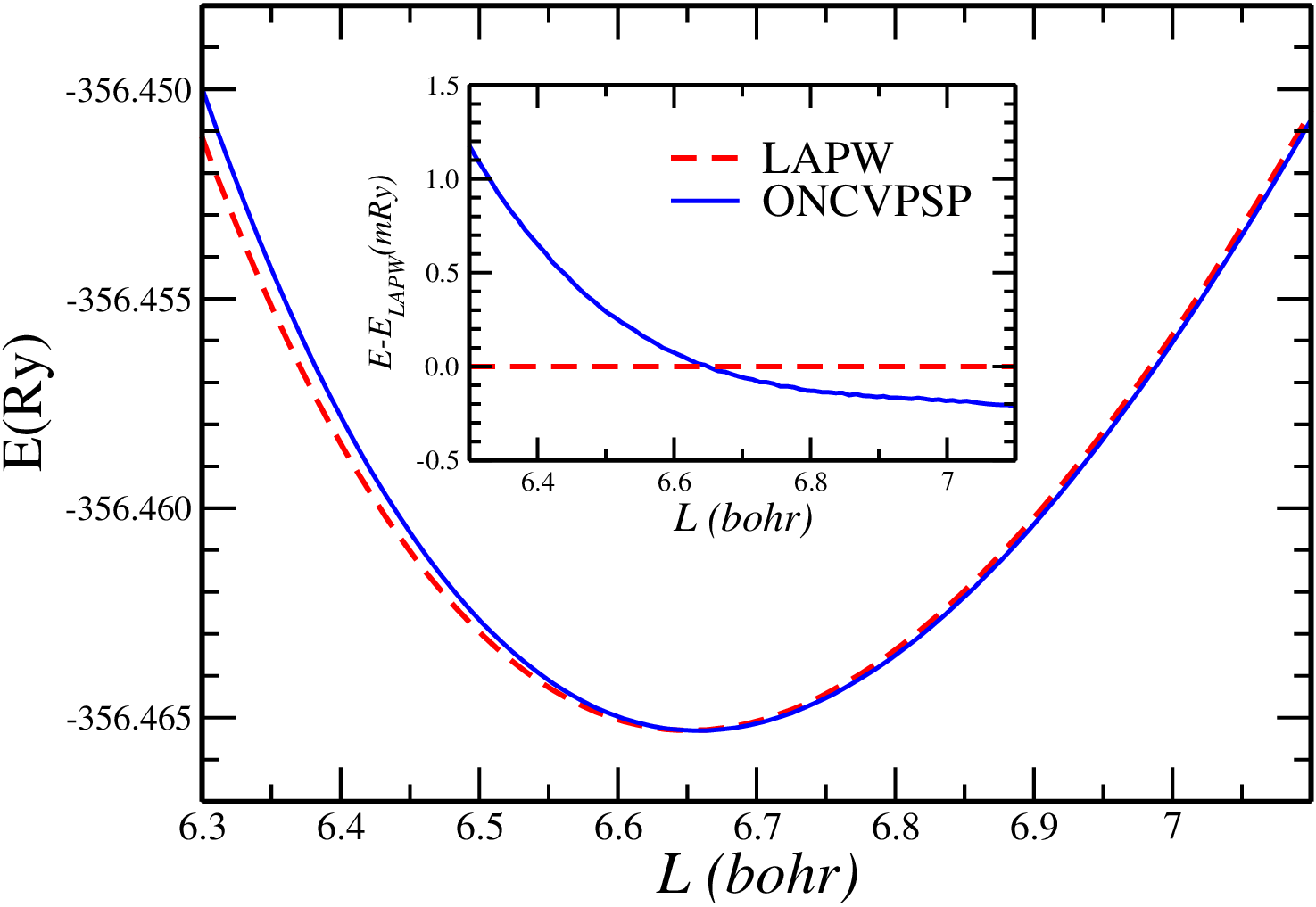}
\caption{\label{Cu_psps} (Color online) 
Cu DFT/LDA EOS comparison of ONCVPSP and LAPW. The EOS are shifted to have the same minimum energy. The inset shows the energy difference versus lattice size. 
}
\end{center}
\end{figure}

For good transferability, a frozen neon-core Cu pseudopotential is required, retaining the $3s^{2}3p^{6}3d^{10}4s^{1}$ states. Single-projector 
NCPPs are challenged in this regard, because the $l=0$ and $l=1$ scattering properties 
near the Fermi energy $E_\mathrm{F}$ depend on
projectors  constructed at much lower energies from the semicore $3s$ and $3p$ states. 
Even the $l=2$ scattering properties near $E_\mathrm{F}$ are difficult, due to the resonant nature of $3d$ scattering. 
To maximize the accuracy, 
very hard single-projector NCPPs must be used, with large  planewave $E_\mathrm{cut}$ $\sim 200$\,Ry.
This is alleviated by the multiple-projector ONCVPSP. Projectors for $l=0$ and $l=1$ can be constructed using both the semicore $3s$ and $3p$ 
and higher-lying valence or virtual $4s$ and $4p$ states. 
Similarly, two reference energies can be used to closely reproduce the all-electron $l=2$ scattering.
We used \mbox{$E_\mathrm{cut}=64$\,Ry} 
and radial cutoffs\cite{ONCVPSP} of 
$r_c=1.60,1.97, 1.97$\,bohr for  $l=0, 1, 2$, respectively. 
The projectors were constructed using the ONCVPSP code,\cite{oncvcode} with the LDA exchange-correlation functional. 
The multiple-projector pseudopotential yields very good agreement with 
 all-electron LAPW results at the DFT level, giving $\Delta$ factor $\sim 1.6$\,meV, as shown in Fig. \ref{Cu_psps}.  
 The non-parallellity error of $\sim 1$\,mRy in the computed EOS  is smaller than the targeted statistical 
 resolution of the QMC calculations  which we discuss next.

\begin{figure}
\begin{center}
\includegraphics[width=0.4\textwidth]{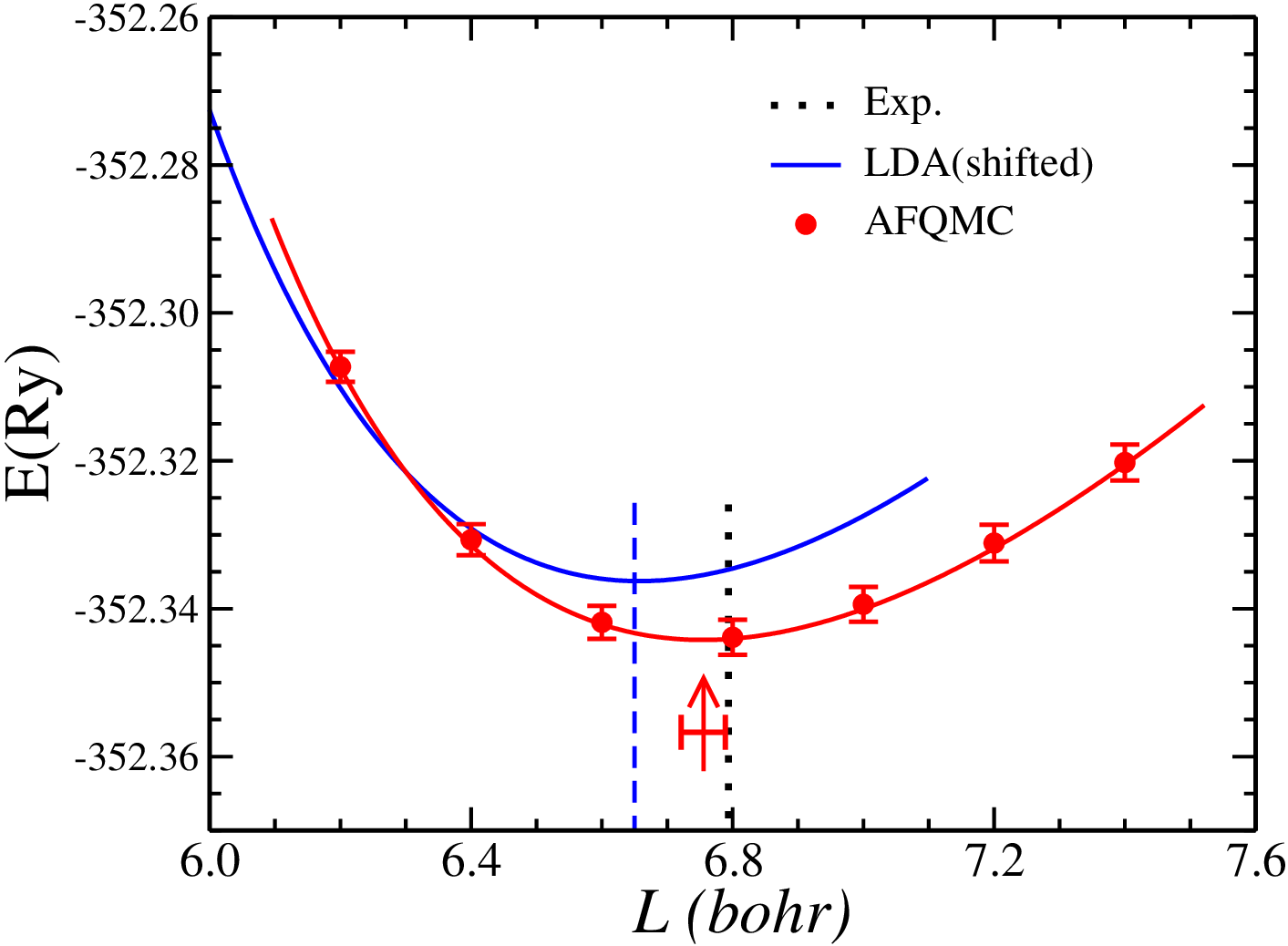}
\caption{\label{Cu_EOS} (Color online) 
Cu  EOS calculated by pw-AFQMC 
using the same ONCVPSP as
in Fig.~\ref{Cu_psps}. 
AFQMC results are shown by filled symbols, with statistical error bars indicated.
For comparison, the DFT EOS with ONCVPSP is reproduced from Fig.~\ref{Cu_psps} (blue solid line, energy shifted). 
The vertical blue dashed and black dotted lines indicate the DFT and experimental equilibrium lattice constants $a_0$, respectively.
The pw-AFQMC calculated $a_0$ 
is indicated by the vertical red arrow with 
horizontal error bar indicating the uncertainty  from a fit to Murnaghan's equation. \cite{MurnaghanEOS}
}
\end{center}
\end{figure}

Figure\,\ref{Cu_EOS} shows the calculated pw-AFQMC Cu EOS. A four-atom cubic simulation cell was used. Because Cu is metallic, 
twist-averaging with a \mbox{$6 \times 6 \times 6$} Monkhorst-Pack (MP) $\mathbf{k}$-point grid \cite{MPgrid} was applied. 
Small random distortions were applied to each of the  $\mathbf{k}$-points to lift band degeneracy in the trial wave function. 
Additionally, post-processing one- and two-body finite-size error corrections \cite{KZKFS,KZKFSspin} were applied. 
The residual finite-size error is not expected to affect the EOS around equilibrium significantly.
This was verified with the following approximate estimate which helps to avoid many computationally costly QMC tests.
Calculations with up to $\mbox{$4 \times 4 \times 4$}$ of primitive unit cell  
were carried out using the LDA+$U$ method. 
The DFT+$U$ method includes a mean-field treatment of on-site 3d electron-electron interactions on the Cu atoms. This effect is absent in standard DFT local and semilocal exchange-correlation functionals, which are based on electron gas calculations. 
Since the choice of $U$ is largely determined by experience and by systematic benchmarking, multiple effective values of $U$, from $0.001$ to $5.0$, are studied in the simulations. 
The same twist-averaging and post-processing finite-size techniques were applied to the LDA+$U$ results. The equilibrium lattice constant and bulk modulus did not change up to the largest test simulation cells. 
The final calculated pw-AFQMC EOS in Fig.\,\ref{Cu_EOS} yields equilibrium lattice constant and bulk modulus, 
\mbox{$a_0=6.76(3)$\,bohr} and \mbox{$B_0=155(13)$\,GPa}, which are in excellent agreement with experimental values 
\mbox{$a_0=6.79$\,bohr}     and \mbox{$B_0=145$\,GPa} (zero-point effects removed) \cite{ExpNaClCu}.
The accuracy of the ONCVPSP compared to all-electron LAPW results at the  
DFT level, without partial core corrections (see Sec.~\ref{sec:Discussion}),
 is thus seen to be a good predictor
of its transferability at the pw-AFQMC many-body level.

\subsection{Sulfur hydride high-$T_c$ high-pressure superconductor:  H$_3$S}

In this section, we present  benchmark pw-AFQMC calculations on two  candidate structures 
for high-temperature, high-pressure  superconductivity in the sulfur hydride system.
Applying the multiple-projector pseudopotentials, we test DFT/GGA predictions of the structural energetics of 
 H$_2$S and H$_3$S by comparison with many-body AFQMC results.

Since Ashcroft proposed that metallic hydrogen should 
exhibit superconductivity with $T_c\sim 270\>{\rm K}$,\cite{PhysRevLett.21.1748} 
there have been many investigations of prospective high-$T_c$ materials incorporating hydrogen, with a recent focus on hydrides, where reduced
metallization pressures are expected. \cite{Ashcroft2004} 
Recent theoretical 
predictions\cite{Li:2014,Duan2014} 
of unusually high 
$T_c$
in sulphur hydrides under high pressure were subsequently supported by experiment. \cite{Drozdov:2014,HSexp,Troyan2016_exp,Einaga2016}
Measurements of resistivity and magnetic susceptibility indicate superconducting temperatures as high as $T_c=203$\,K 
at pressures  $\sim 150$\,GPa;  \cite{HSexp} this was attributed to the  {\it $Im\bar{3}m$} H$_3$S phase.
A novel experiment reported Meissner effect measurements that qualitatively confirmed the finding. \cite{Troyan2016_exp}
Subsequent DFT-based calculations have led to similar conclusions regarding the central role of electron-phonon coupling in driving
the superconducting transition.
\cite{MazinHnS,DuanPRB2015,NRLgroup2015,Papacon2015,ErreaPRL,Komelj2015}
These calculations support the view that the sulfur hydrides are  
conventional superconductors,
which are well described by Bardeen-Cooper-Schrieffer (BCS) theory \cite{BCS} with strong electron-phonon coupling leading to 
high $T_c$. This is unlike the previously known high-$T_c$ cuprate and iron-based  superconductors, where 
strong  
electron-electron interactions are believed to play a key role, although the superconducting
mechanism has not yet been established.
With a $T_c \sim 203$\,K,\cite{HSexp} hydrogen sulfide is one of the highest temperature superconductors on record, although extremely high pressures
are required. Their discovery has re-energized the search for new superconductors  in hydrogen-based and related materials.

Little is known experimentally regarding the high-pressure stability of hydrogen sulfide compounds. There has therefore
been a strong reliance on standard DFT calculations, which have examined the high pressure phase stabilities and structures of H$_\mathrm{n}$S. \cite{MazinHnS,DuanPRB2015,ErreaPRL}. The H$_3$S {\it $Im\bar{3}m$} structure  (space group No.~$229$) has been
a leading candidate for the stoichiometry that leads to highest $T_c$.  Other stoichiometries like H$_2$S are predicted to have 
competitive but less favorable enthalpies. It is important, therefore, to test these predictions with accurate many-body calculations.
Here, we focus on candidate structures for two compositions, H$_2$S and H$_3$S, and compare pw-AFQMC results of their structural energetics 
with DFT/GGA predictions.

\begin{figure}
\begin{center}
\includegraphics[width=0.4\textwidth]{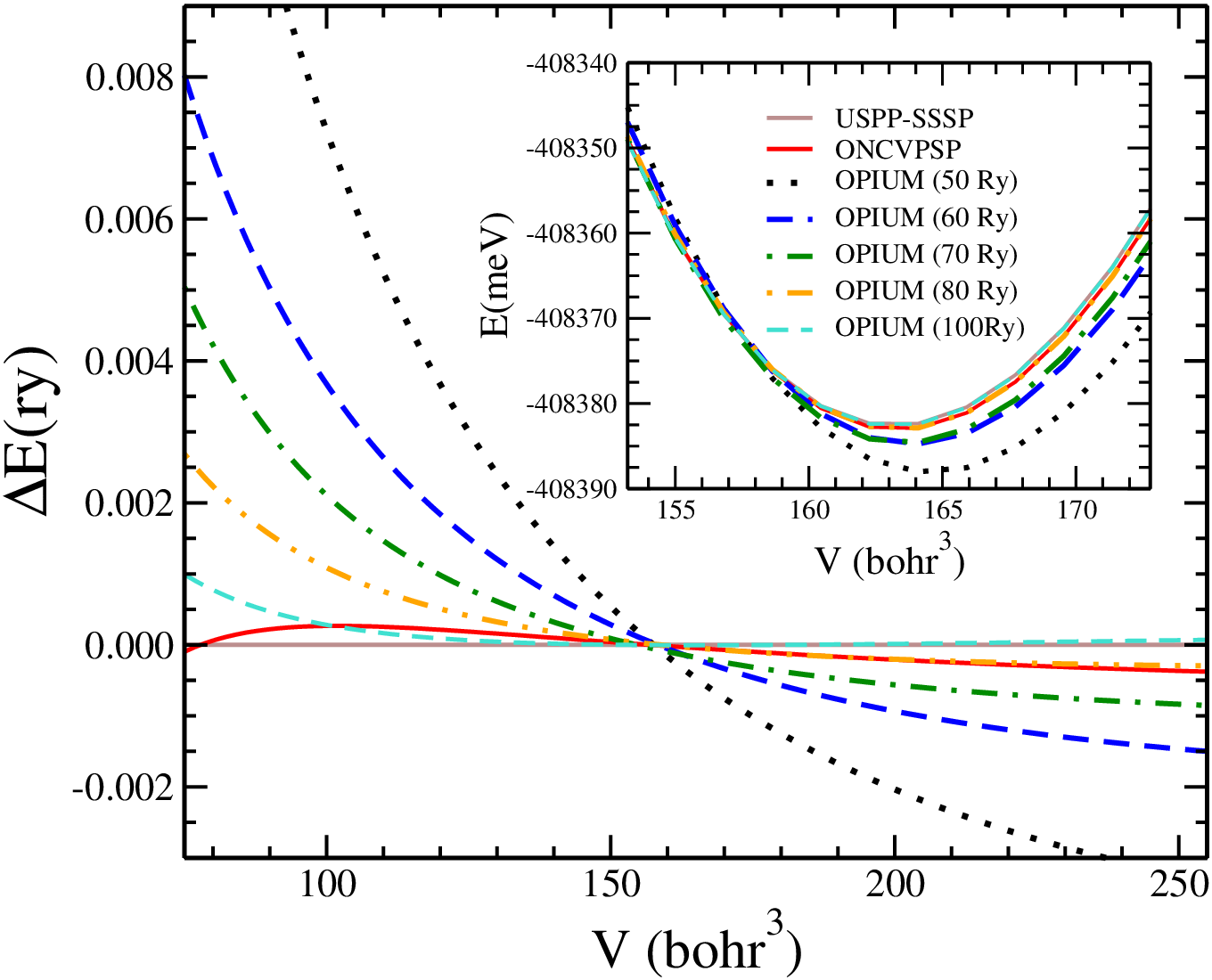}
\caption{\label{H3S_EOS} (Color online) 
H$_3$S ($Im\bar{3}m$) EOS calculated with ONCVPSP, single-projector NCPP, and USPP pseudopotentials, 
plotted as \mbox{$\Delta E(V)=E(V)-E_\mathrm{USPP}$} (minimum energies aligned).
OPIUM generated NCPP EOS are shown for a range of $E_\mathrm{cut}$.
The inset shows the the actual EOS. 
}
\end{center}
\end{figure}

The ONCVPSPs of H and S were generated with  \mbox{$E_\mathrm{cut}=50$\,Ry}. The $l=2$ projectors for S used unbound scattering states.\cite{USPP,ONCVPSP} 
Figure~\ref{H3S_EOS} compares calculated ONCVPSP EOS with ultrasoft (USPP) and single-projector NCPP pseudopotential calculations, 
using the DFT GGA/PBE xc functional. The NCPPs were generated with the OPIUM\cite{opium} package for several values of  \mbox{$E_\mathrm{cut}$}, and USPP ``Standard Solid State Pseudopotentials" (USPP-SSSP) are adopted.\cite{DFT-Reproducibility}
ONCVPSP is in excellent agreement with USPP over a wide volume range 
($\sim 75 - 255~\mathrm{bohr}^3$). The difference is less than 0.5\,mRy per formula unit. 
Using USPP-SSSP as the reference, $\Delta$ is $0.6$\,meV for ONCVPSP over the typical choice of  $\Delta V$, 
$\pm6\%$ around the equilibrium volume of  $V_0 \simeq163~\mathrm{bohr}^3$. The volume range in Fig.~\ref{H3S_EOS}, of $75\sim255~\mathrm{bohr}^3$, 
is much wider, covering a \mbox{$\pm50\%$} span and including the superconducting
 volume near $90~\mathrm{bohr}^3$ at transition pressure 200\,GPa. 
For this volume range, the $\Delta$ is  $3.0$\,meV  for ONCVPSP.
By comparison, the single-projector NCPPs 
have $\Delta$ values of $68.1$, $34.6$, $20.7$, $10.6$, and $3.4$\,meV, for \mbox{$E_\mathrm{cut}$} values
of  50, 60, 70, 80, and 100\,Ry, respectively. 
To achieve comparable accuracy with the ONCVPSP, 
the NCPP requires a \mbox{$E_\mathrm{cut}=100$}\,Ry, which gives a nearly three times larger planewave basis.

\begin{table}
\begin{center}
\begin{ruledtabular}
\begin{tabular}{ l  l  c  c }
                     &               & H$_2$S   & H$_3$S     \\
\hline
                     &               & \mbox{$\mathrm{E}_{64}-\mathrm{E}_{26}$}  & \mbox{$\mathrm{E}_{229}-\mathrm{E}_{66}$} \\
                     &               &  (eV/atom)    & (eV/atom) \\
\hline
\multicolumn{2}{l}{AFQMC}            & -0.086(7) & 0.111(5) \\
\hline
\multirow{3}{*}{LDA} &  ONCVPSP      & -0.082    &  0.058   \\
                     & OPIUM(100Ry)  & -0.083    &  0.056 \\
                     &  USPP         & -0.084    &  0.056 \\
\hline
\multirow{3}{*}{PBE} &  ONCVPSP      & -0.086    &  0.082 \\
                     & OPIUM(100Ry)  & -0.088    &  0.080 \\
                     &  USPP         & -0.086    &  0.080  \\
\hline
\multirow{3}{*}{PBEsol} &  ONCVPSP   & -0.083    &  0.060   \\
                     & OPIUM(100Ry)  & -0.084    &  0.058  \\
                     &  USPP         & -0.084    &  0.059  \\
\hline
PBE0                 &  ONCVPSP      & -0.082    &  0.077   \\
\end{tabular}
\end{ruledtabular}
\end{center}
\caption{\label{Tab_1} 
Calculated pw-AFQMC structural energy difference for H$_2$S and H$_3$S,  using ONCVPSP, 
compared to DFT-based calculations for four functionals and three
pseudopotentials.  For each of the four crystal structures, the fully relaxed \mbox{$P=200~\mathrm{GPa}$} structure from ONCVPSP-PBE was used for
the pw-AFQMC calculations and for 
the other DFT functionals and pseudopotentials. 
}
\end{table}

\begin{figure}
\begin{center}
\includegraphics[width=0.4\textwidth]{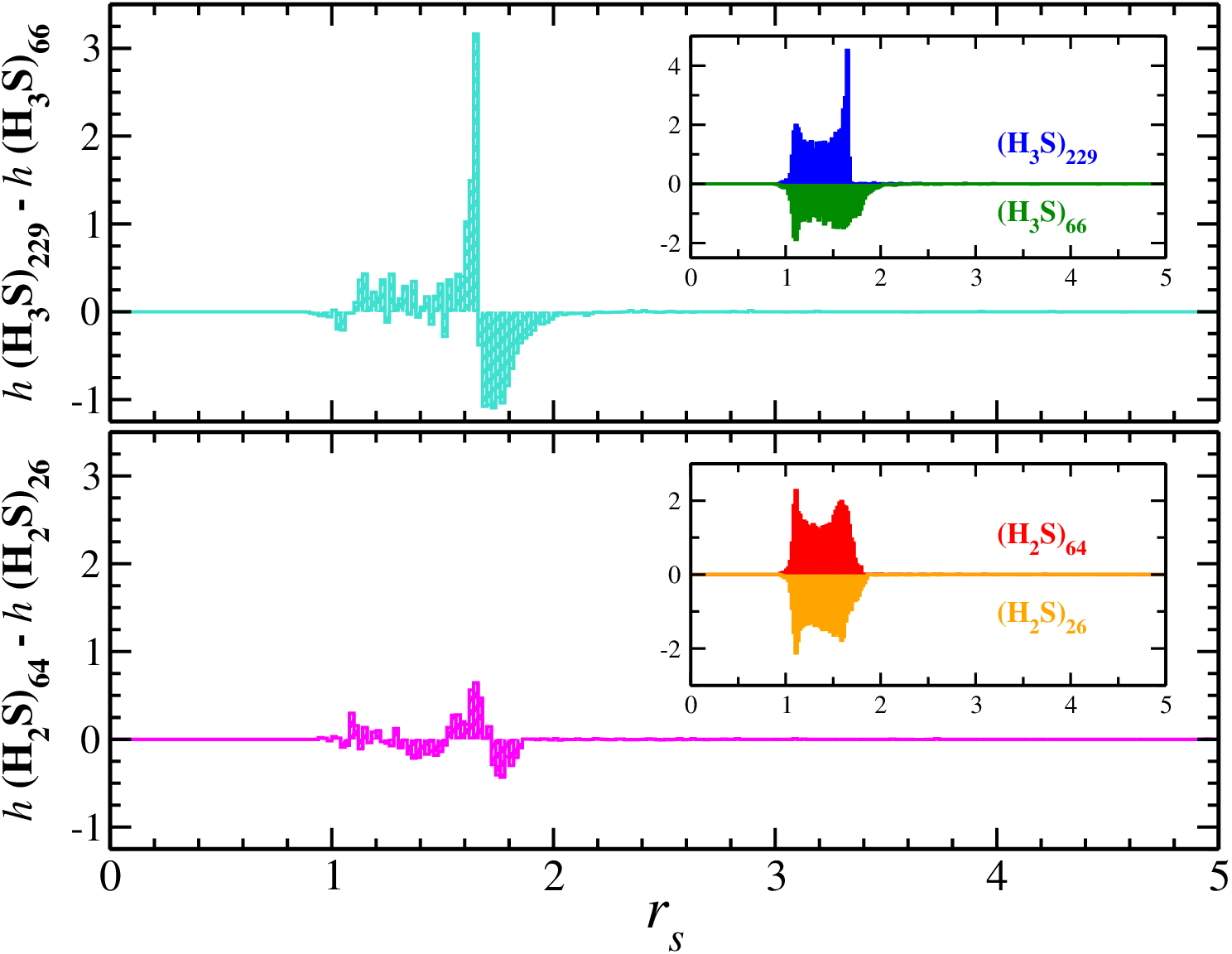}
\caption{\label{H3S_H2S_Rs} (Color online) 
Electron density distributions in H$_{3}$S (top panel) and H$_{2}$S (bottom panel) as a function of  $r_s$, computed from DFT/PBE with ONCVPSP. 
The main plots show the difference between the two space group structures for each composition, while the 
insets show the actual distributions of the two structures, with one shown as negative.
$\Delta r_s$=0.02 is chosen as the size of histogram bin.
}
\end{center}
\end{figure}

For the pw-AFQMC benchmarks, we selected \mbox{200 GPa} structures for two compositions, guided by 
the DFT/PBE results of Mazin {et al.}\cite{MazinHnS} 
For H$_2$S, the space group structures $Pmc2_1$ (\#26) and $Cmca$ (\# 64) were selected, both with 12-atom primitive cells. 
For H$_3$S, the space group structures $Cccm$ (\#66) and  $Im\bar{3}m$ (\#229) were selected, with 32- and 4-atom primitive cells, 
respectively. 
We first calculated DFT-based energy differences  \mbox{$\mathrm{E}_{64}-\mathrm{E}_{26}$} for  H$_2$S and
\mbox{$\mathrm{E}_{229}-\mathrm{E}_{66}$} for  H$_3$S. 
Our \mbox{ONCVPSP} DFT/PBE calculations are in very good agreement with the results in Ref.~\onlinecite{MazinHnS}. 
DFT-based results are shown in
Table \ref{Tab_1} for combinations of three pseudopotentials (ONCVPSP, NCPP-100Ry, and USPP) and four DFT exchange-correlation functionals: LDA, PBE, PBEsol, and the hybrid PBE0 method. 
(Note that the sign of the energy differences does not reflect the relative structural stabilities. For example, the calculated 
 H$_3$S DFT/PBE enthalpy is actually lowest,\cite{MazinHnS} making it the most stable structure at \mbox{$P=200$\,GPa}.)
To facilitate comparisons, the fully relaxed \mbox{200 GPa} crystal structure from ONCVPSP-PBE was used
for the other functionals and pseudopotentials and for the pw-AFQMC calculations. 
For H$_2$S,  \mbox{$\mathrm{E}_{64}-\mathrm{E}_{26}$} is nearly independent of the choice of DFT functional, while
for H$_3$S, \mbox{$\mathrm{E}_{229}-\mathrm{E}_{66}$} varies between 0.056 to 0.082 eV/atom. 

For the pw-AFQMC calculations, 24-atom simulation cells were used for H$_2$S,  doubling the 
size of the primitive unit cell in each structure. 
For H$_3$S, 32-atom simulation cells were used \mbox{($2\times2\times2$ for $Im\bar{3}m$)}.
Twist-averaged boundary conditions
with a $4\times4\times4$ MP grid 
were used. One- and two-body finite-size corrections \cite{KZKFS,KZKFSspin} were then applied to the many-body results. The pw-AFQMC energy differences are also shown in Table \ref{Tab_1}. 
The pw-AFQMC H$_3$S energy difference, 0.111(5)~eV/atom, is nearly twice that given by the LDA and PBEsol, 
and about 50\% larger than those from PBE and PBE0, while in H$_2$S the DFT-based calculations are identical with pw-AFQMC to within its statistical
uncertainty. 

To understand how the better agreement in H$_2$S arises compared to H$_3$S, we 
investigated the electron-density distributions for each composition.
The result is illustrated in Fig.~\ref{H3S_H2S_Rs}, which plots the  densities  
calculated from ONCVPSP DFT/PBE 
for the four structures on the real-space FFT grid.  
In both H$_3$S and H$_2$S, the distribution is largely concentrated in the high-density region $r_s = 1~\mathrm{to}~2$. 
The H$_3$S composition structures, however, show larger differences, especially in the range
 $r_s=1.6~\mathrm{to}~2.0$,
 than for the two H$_2$S structures.
This provides a possible explanation of the better agreement of the different DFT functionals for H$_2$S than for H$_3$S. Similarly,
it  indicates that there will be better cancellation of electron correlation effects in H$_2$S, resulting
in better agreement between DFT and 
pw-AFQMC many-body results.

The pw-AFQMC benchmarks in Table \ref{Tab_1} show that DFT-based predictions are semi-quantitatively correct.
The DFT predictions
 of H$_2$S and H$_3$S enthalpy differences could be off by the order of 30\,meV and 50\,meV 
 in PBE and PBEsol, respectively. 
 However, the stabilities  
 are dominated by 
 independent-electron contributions to the enthalpy, which are significantly larger than these differences.
 This suggests that the predictions on phase stabilities and structures from recent DFT studies are likely reasonable.

\section{Discussion}
\label{sec:Discussion}

The applications above show that the use of multiple-projector ONCVPSP 
can greatly reduce the planewave basis size in pw-AFQMC many-body calculations, while maintaining or improving accuracy compared to single-projector NCPPs.
ONCVPSP
 uses two projectors per partial wave in our applications, which maintains fidelity to scattering properties 
at reduced $E_\mathrm{cut}$. 
As discussed in Section~\ref{sec:GSprojection}, this results in significant reductions of  the computational cost, both by reducing the computing time for each
step in the random walk and, at the same time, increasing the QMC efficiency  
because of a smaller number of auxiliary fields in Eq.~(\ref{eq:HS-xform0}). 
For example, accurate results were obtained in fcc Cu with $E_\mathrm{cut}=64$\,Ry, in contrast 
to an estimated value of $E_\mathrm{cut}\sim 200$\,Ry with NCPP.

It is important to note, however, that 
improvement in performance in DFT calculations  by the ONCVPSP over single-projector NCPP 
does not always correlate with improvement in QMC. 
There are fundamental differences in the role of DFT-generated pseudopotentials
when applied in a many-body context, versus in DFT. Clearly, when core-valence  (or core-core) correlation effects are non-negligible,  the use of 
pseudopotentials generated from an independent-electron approach
can incur errors in many-body calculations. 
This is not the case for the systems treated in this paper. 
For example, in NaCl, 
small-core pseudopotentials are taken to pseudize the $2s$, $2p$ states
in both DFT and AFQMC.
In DFT, partial-core effects were negligible as shown by the good agreement
with LAPW in Fig.~1. Similarly in AFQMC, excellent agreement is found with experiment in Fig.~2.

A case that illustrates the difference is 
bulk Si, where the Ne-core NCPP causes a pseudopotential error both in DFT and AFQMC. 
In DFT, this can be remedied using a partial-core correction, which is not available in AFQMC.
One way to remove the Ne-core error in AFQMC is with the frozen-core (FC) approximation. \cite{AFQMCFC,AFQMCdownfolding}
A He-core pseudopotential is used to generate DFT $2s$ and $2p$ orbitals in the crystalline solid environment.
After a unitary rotation to the Kohn-Sham basis, the $2s$ and $2p$ orbitals are frozen \cite{AFQMCdownfolding}.
 The corresponding FC
Hamiltonian, which incorporates an effective Ne-core pseudopotential, 
was shown to yield excellent results 
\cite{AFQMCdownfolding} (in which the NCPP for the He-core
pseudopotential used in the DFT calculations to generate the FC orbitals and Kohn-Sham basis
had an extremely high cutoff, $\sim$ 600\,Ry). 
To further study the implicit treatment of core-valence interactions in the FC approximation,
we repeated this procedure with a much softer He-core ONCVPSP ($\sim$ 64\,Ry).

At the DFT level, this ONCVPSP  
works as well as the 600\,Ry NCPP.
They both capture (treating 12 electrons/Si) the core-valence corrections and yield excellent agreement 
with all-electron LAPW and with partial-core-corrected Ne-core pseudopotential calculations.
However, the corresponding FC AFQMC calculation is not improved, and a pseudopotential error 
is seen as in the Ne-core calculation.  This is because the softer  
He-core ONCVPSP has 
large pseudizing radius cut-offs which affect the resulting  $2s$ and $2p$ orbitals.
In DFT, such errors can be partially recovered because the densities are properly compensated for. In QMC, 
when the 
less accurate $2s$ and $2p$ orbitals are frozen (treated at the Hartree-Fock level), the errors propagate 
into the FC many-body 
Hamiltonian that the QMC treats, which cannot be corrected.

 A good indicator of the accuracy of ONCVPSPs in many-body calculations is thus good core-valence separation and good DFT performance of ONCVPSP (without partial-core corrections) compared 
 to all-electron calculations.  
 The improved ONCVPSP scattering properties and transferability then allow smaller values of $E_\mathrm{cut}$,
 which can significantly reduce the many-body computing cost while retaining high accuracy.
 In intermediate cases, such as in Si, when partial-core corrections are necessary in DFT calculations, 
 more care is required.

\section{Summary}
\label{sec:Summary}

We have successfully implemented the multiple-projector ONCVPSP into the many-body pw-AFQMC method. The accuracy is demonstrated by calculations of bulk properties of NaCl and the more strongly correlated fcc Cu. With this technique, we also benchmarked the structure transition energy barriers in the recently discovered high-temperature superconductor sulfur hydride systems. 
In these systems,
modest electron-electron correlation and large cancellation effects are seen in the energies between different structures, and
we find that the estimations from DFT are in reasonable agreement with the many-body AFQMC results.
The implementation of multi-projectors pseudopotential allows pw-AFQMC to treat systems 
with smaller  pseudopotential errors and at significantly lower planewave energy cut-offs, 
 and hence to reach larger and more complicated systems.

\acknowledgements

This work is supported by DOE (DE-SC0001303), NSF (DMR-1409510), and ONR (N000141211042). An award of computer time was provided by the Innovative and Novel Computational Impact on Theory and Experiment (INCITE) program, using resources of the Oak Ridge Leadership Computing Facility at the Oak Ridge National Laboratory, which is supported by the Office of Science of the U.S. Department of Energy under Contract No. DE-AC05-00OR22725. We also acknowledge computing support from the computational facilities at the College of William and Mary.

\bibliography{ONCVPSP}

\end{document}